\documentclass[11pt,twoside]{article}

\usepackage{asp2006}
\usepackage{epsf}
\usepackage{epsfig}
\usepackage{lscape}
\usepackage{natbib}

\begin{document}
\title{Radial Velocity Variable sdO/Bs from SPY - Preliminary Orbits of Three New Short Period Binaries}% Fill in title
\author{S. Geier$^1$, U. Heber$^1$, and R. Napiwotzki$^2$}   %%% Fill in author names
\affil{$^1$ Dr.~Remeis--Sternwarte, Institute for Astronomy, University Erlangen-Nuremberg, Sternwartstr. 7, 96049 Bamberg, Germany}
\affil{$^2$ Centre of Astrophysics Research, University of Hertfordshire, College Lane, Hatfield AL10 9AB, UK}

    %%% Fill in author affiliations

\begin{abstract} %%% Abstract to run on from here.
We present new results from follow-up time-resolved spectroscopy of
radial velocity variable hot subdwarfs from the SPY project. Medium
resolution spectra were taken at the ESO NTT. Preliminary orbital
solutions are presented for three single-lined sdO/B binaries. The
orbital periods of all binaries are short and the companions are most
likely white dwarfs or late main-sequence stars. Follow-up
time-resolved 
spectroscopy is necessary to measure the orbital parameters
with higher accuracy. Light curves should be taken in order to search
for reflection effects and or eclipses.
\end{abstract}

\section{Introduction}   

The ESO Supernovae Ia Progenitor Survey (SPY) aimed to find
double-degenerate 
binaries in close orbits \citep{piii2_napiwotzki3}. The merger of
two white dwarfs is considered to be one of the processes, which could
lead to a supernovae Type Ia (SN Ia) explosion. A large spectroscopic
survey of 1\,000 white dwarf (WD) candidates was undertaken using the
UVES spectrograph at the 8-m ESO VLT UT2 (Kueyen) to search
for radial velocity (RV) variable WDs. About 140 hot subdwarfs of
various types were included. Hot subdwarf B stars (sdBs) are
considered to be core helium-burning stars with very thin hydrogen
envelopes which are situated on the Extreme Horizontal Branch
(EHB). The formation of sdBs is still unclear. Different formation
channels have been discussed \citep{piii2_han}. As it turned out, a large
fraction of the sdB stars are members of short-period binaries
\citep{piii2_maxted2, piii2_napiwotzki4}. For these systems common-envelope
ejection is the most probable formation channel. Hot subdwarf binaries
therefore provide a suitable population to study this very important,
but still poorly understood phase of stellar evolution. Most
companions of sdBs in such systems are white dwarfs (WDs) or late type
main-sequence (MS) stars \citep{piii2_morales, piii2_edelmann}. Close binary
sdB+WD systems, which exceed the Chandrasekhar mass, turned out to be
good candidates for double-degenerate progenitors of SN\,Ia
\citep{piii2_maxted1, piii2_geier}.  Measuring the orbits of close binary sdBs is
therefore necessary to study common-envelope evolution and a
prerequisite for more detailed studies to constrain the system
parameters.

\section{Observations and Data Analysis}

The programme stars were observed at least twice in the course of the
 SPY project with the high-resolution echelle spectrograph UVES at the
 ESO\,VLT (2000--2002). Follow-up observations were undertaken with
 the medium resolution spectrograph EMMI at the ESO NTT in June
 2007. Reduction was done with the ESO--MIDAS package. The radial
 velocities (RV) were measured by fitting a set of mathematical
 functions (Gaussians, Lorentzians and polynomials) to the hydrogen
 Balmer lines using the FITSB2 routine \citep{piii2_napiwotzki2}. Sine
 curves were fitted to the RV data using single-value
 decomposition, a \(\chi^{2}\)
 minimising method, and the power spectrum was generated.

\section{Orbital Parameters, Minimum Companion Masses and Nature of the Unseen Companions}
\begin{table}[t!]
\caption{Orbital periods, radial velocity semi-amplitudes, mass
functions and minimum masses of the companions (assuming $M_{\rm
sdB}=0.47\,M_{\rm \odot}$). The orbital parameters are similar to
known close binary sdBs with short periods \citep{piii2_morales}.}
\centering
\smallskip
\begin{tabular}{lllllll}
\hline
\noalign{\smallskip}
 & $P$ & $K$ & $f(M)$ & $M_{\rm 2min}$ & \\ 
& [d] & [${\rm km\,s^{-1}}$] & [$M_{\rm \odot}$] & [$M_{\rm \odot}$] &\\
\noalign{\smallskip}
\hline
\noalign{\smallskip}
WD\,0107$-$342 & $0.375\pm0.05$ & $127\pm2$ & $0.08$ & $0.39$ & sdB+MS/WD\\
HE\,1415$-$0309 & $0.163\pm0.03$ & $228\pm8$ & $0.20$ & $0.62$ & sdB+WD\\
HE\,1423$-$0119 & $0.197\pm0.01$ & $44\pm11$ & $0.001$ & $0.08$ & sdO+MS/WD\\
\noalign{\smallskip}
\hline
\end{tabular}
\end{table}

\begin{figure}[tp!]
 \plottwo{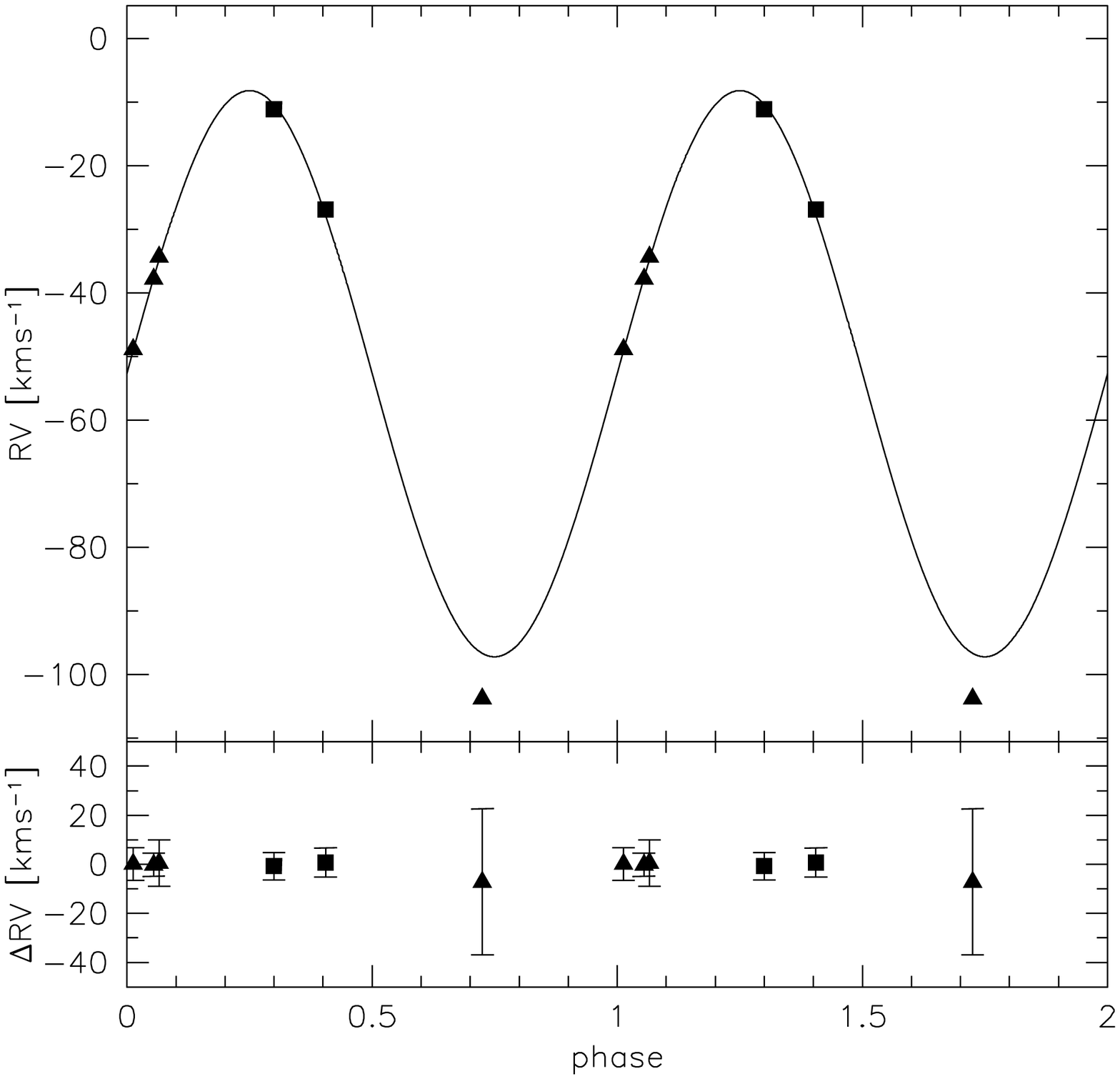}{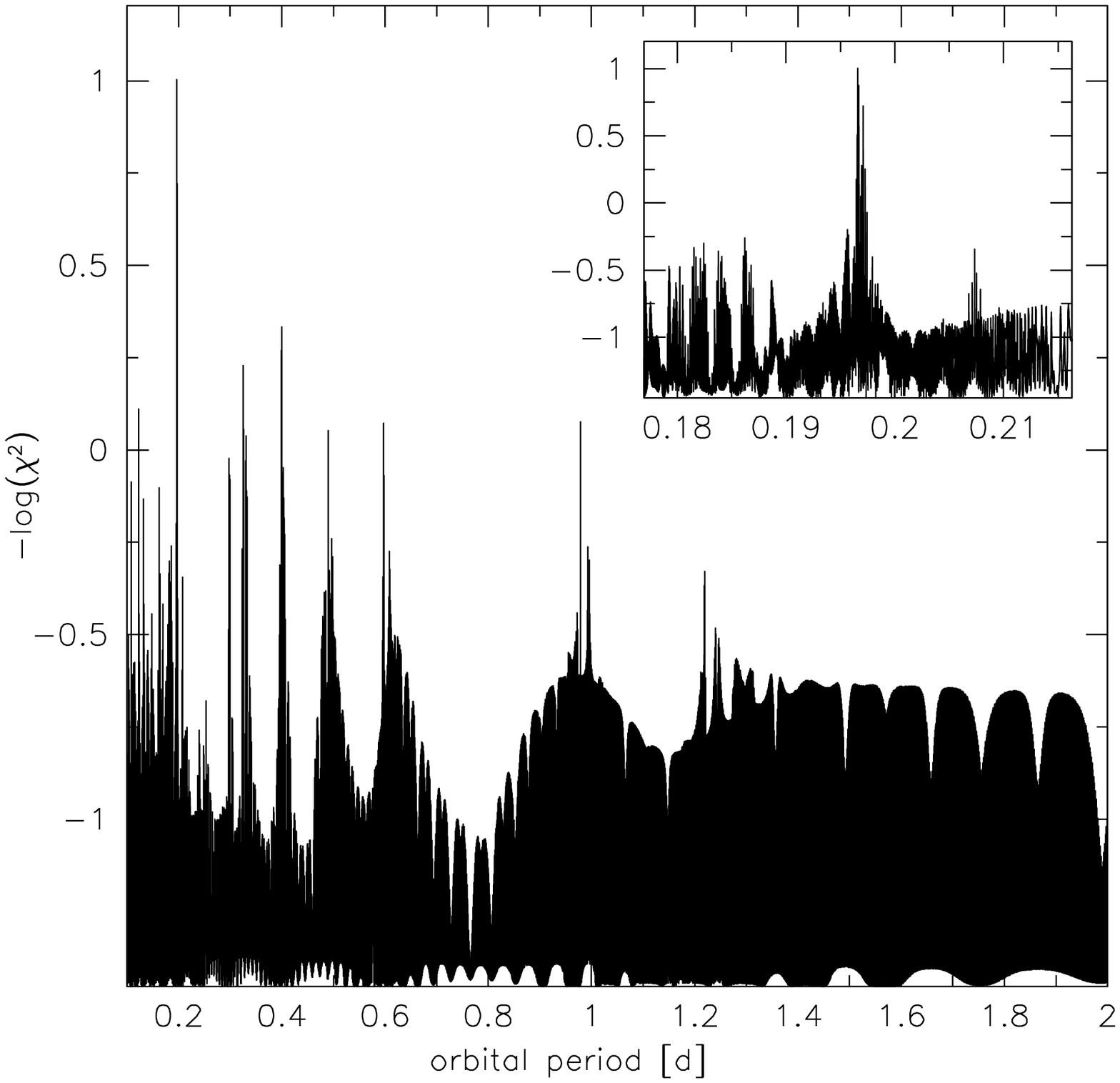}
 \plottwo{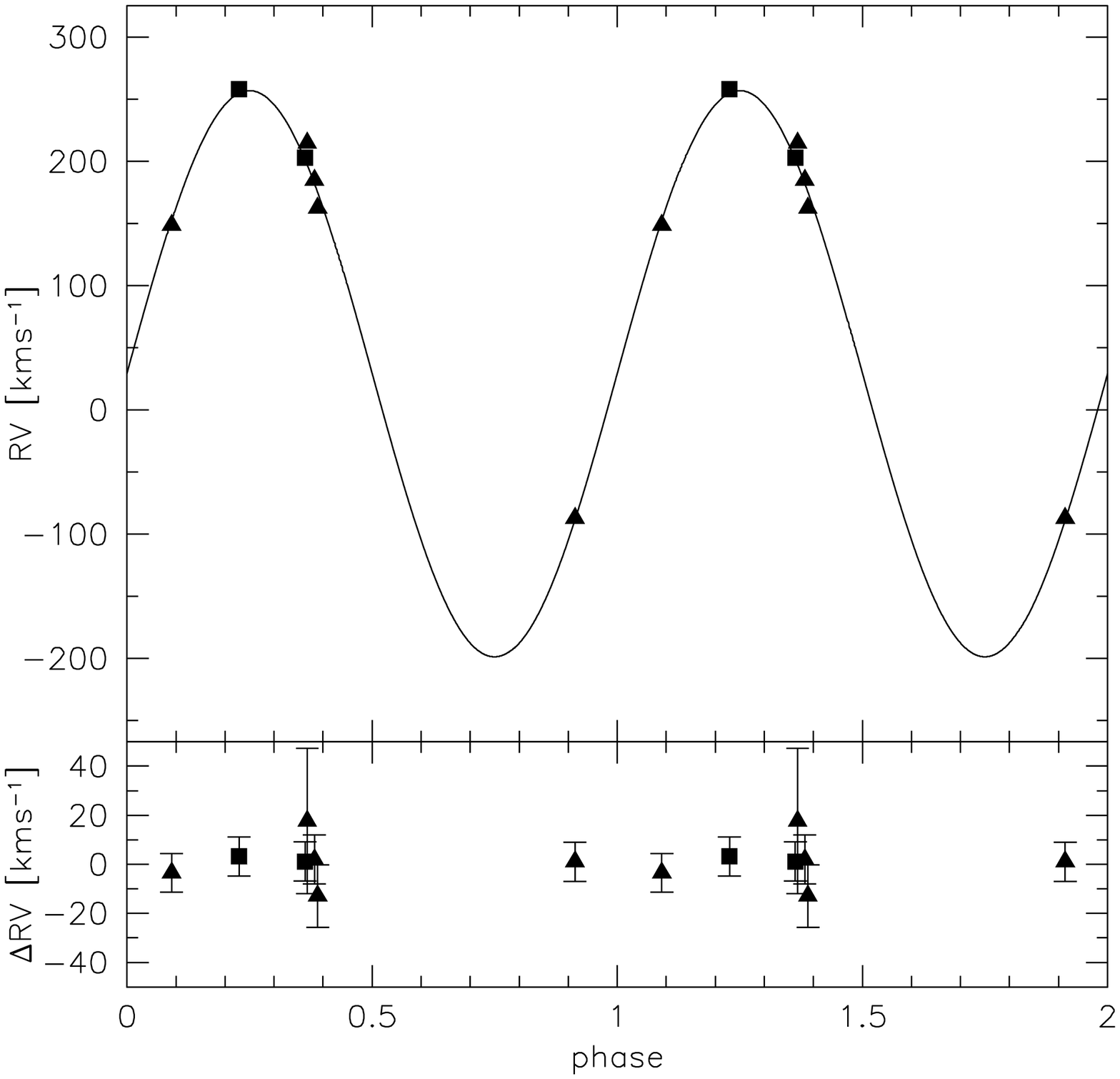}{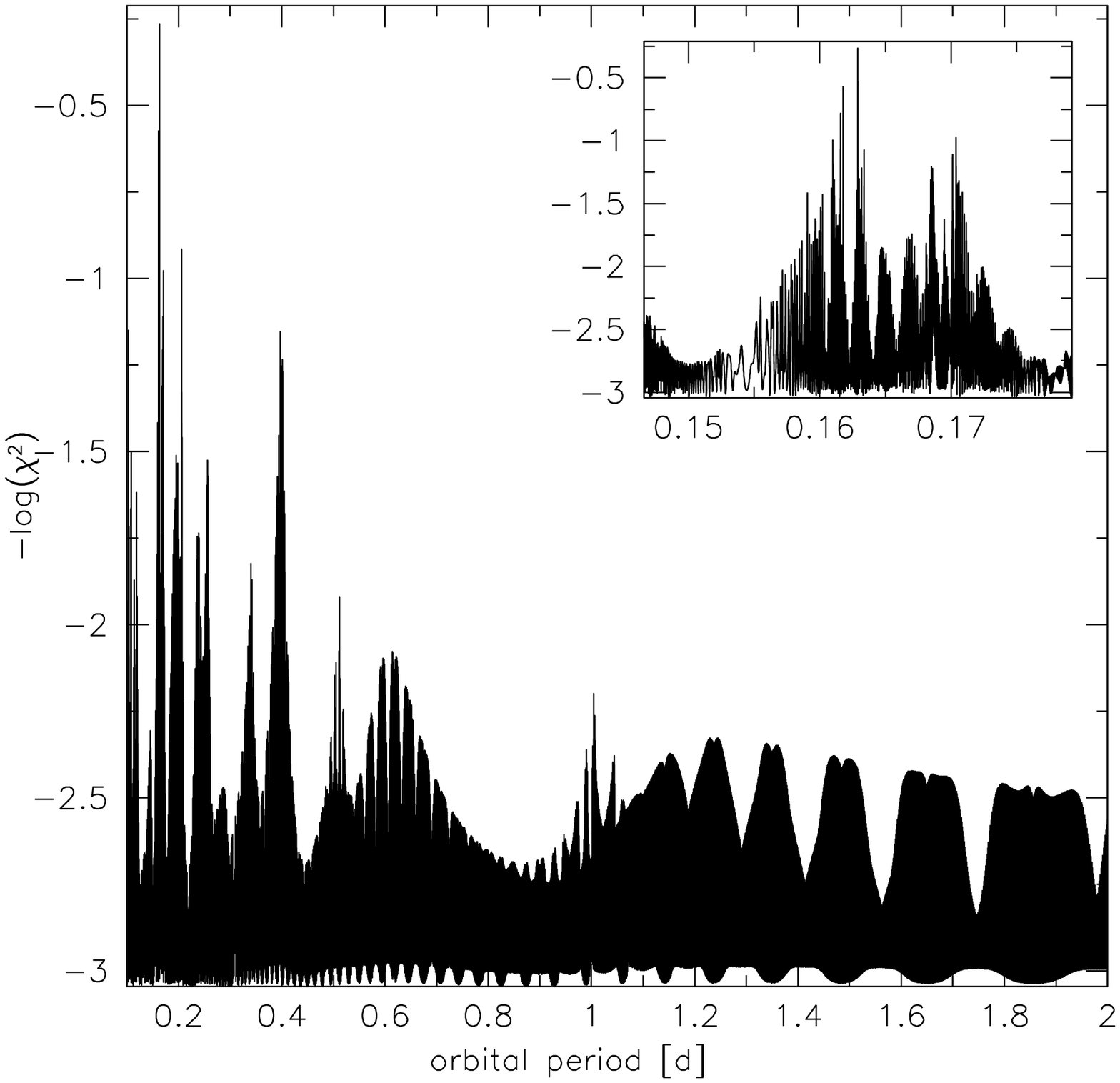}
 \plottwo{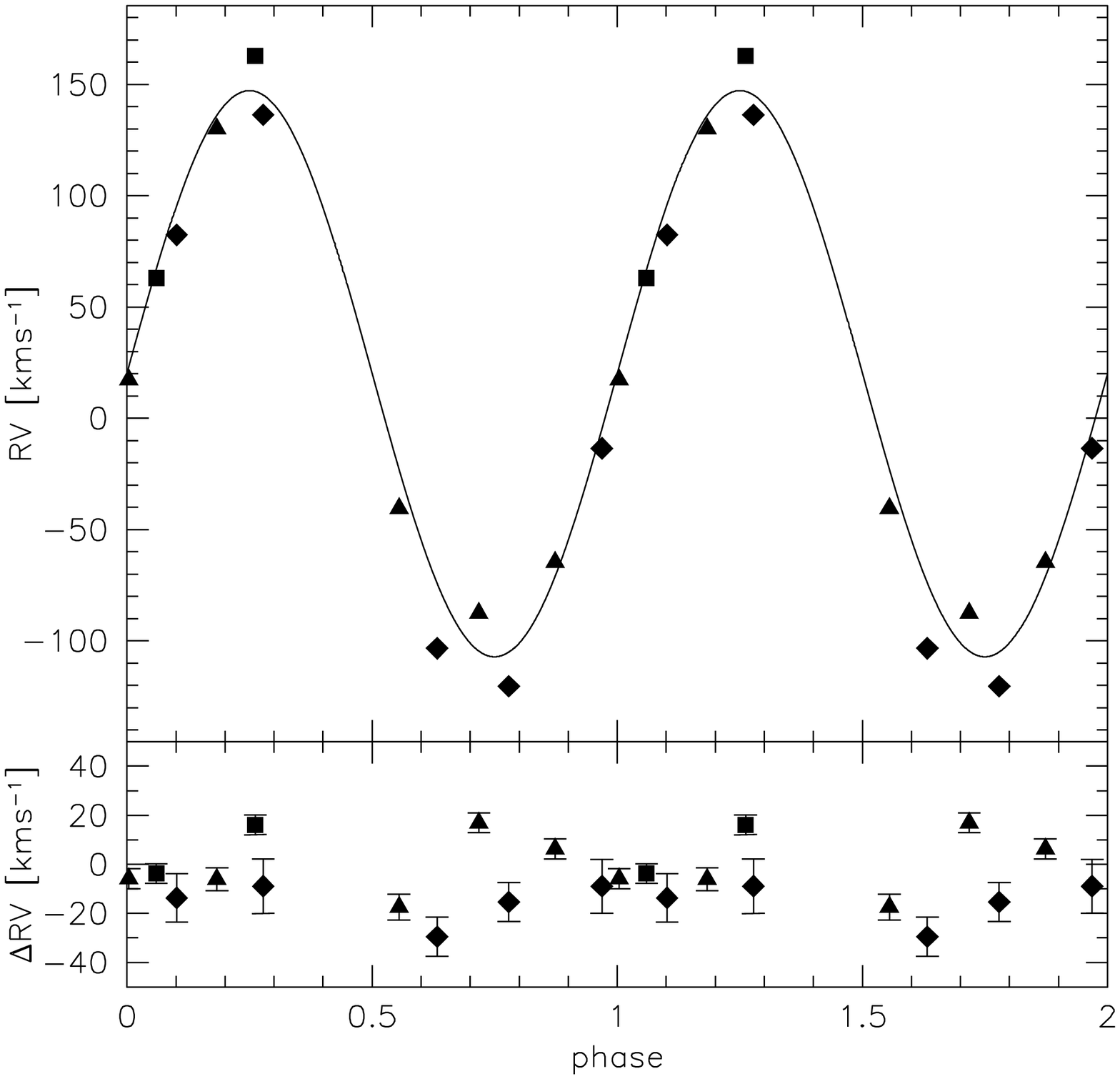}{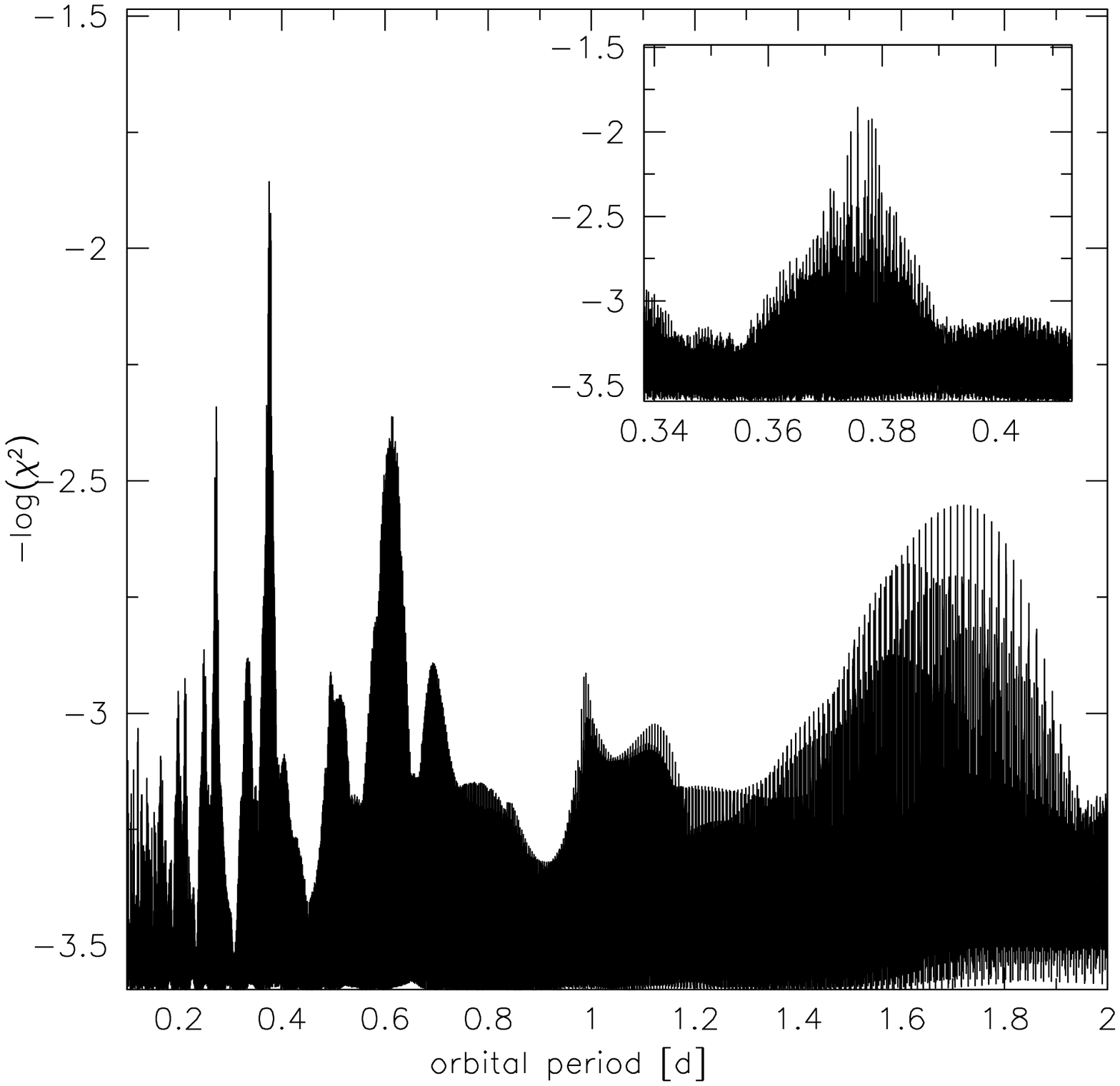}
 \caption{{\it Left panel:} Subdwarf radial velocities against orbital
 phase for HE\,1423--0119 (top), HE\,1415--0309 (middle), and
 WD\,0107--342 (bottom). The
 residuals and errors are plotted below for VLT (squares), NTT run 2004 (diamonds) 
 and NTT run 2007 (triangles) velocities. {\it Right panel:} Power spectra
 $-\log\chi^{2}$ of the best sine fits are plotted against orbital
 period. Best solutions are expanded in the inset boxes.}
\end{figure}

All three binaries presented here are single-lined systems. This
yields an upper limit for the mass of a late main sequence companion
of $0.45\,M_{\rm \odot}$, which should otherwise be visible in the
spectrum. 
HE\,1415$-$0309 has a very short period and high radial
velocity amplitude. The minimum mass of the secondary exceeds
$0.45\,M_{\rm \odot}$ and therefore excludes a main sequence
companion. It is very near the average mass of DA white dwarfs
\citep[$0.59\,M_{\rm \odot}$;][]{piii2_koester}. A high inclination is
therefore probable and the system could be eclipsing. HE\,1415$-$0309
is likely to be an sdB+WD binary, but a heavy compact companion
cannot be ruled out. HE\,1423$-$0119 is a hydrogen rich sdO star in a
close orbit. Due to its low minimum mass the companion may be a
late-type 
MS star. In this case, a reflection effect may be detectable in
the lightcurve. The minimum companion mass of WD\,0107$-$342 is
compatible with a WD as well as a late MS star. 

Assuming orbital
synchronization of the binary and measuring the subdwarfs projected 
rotational velocity $v_{\rm rot}\sin{i}$ from high resolution spectra, 
it is possible to derive the inclination of the system as well as the companion mass. 
In this case the companion is most likely a heavy
white dwarf and WD\,0107$-$342 becomes a viable candidate for a SN Ia
explosion (see Geier et al. these proceedings).

\section{Discussion}

The results presented here should be considered to be preliminary. As can
be seen from the RV curves, more data points are needed to get
better phase coverage, in particular for HE\,1415$-$0309 and
HE\,1423$-$0119. In the course of the SPY survey 18 radial velocity
variable subdwarfs were discovered. The orbital parameters of 12 close
binary subdwarfs have already been determined \citep{piii2_napiwotzki1,
piii2_napiwotzki4}. The goal of our project is to measure the orbital
parameters of all RV variable subdwarfs from the SPY
survey. Additional NTT time was granted for Period 80 in January 2008.

To constrain the nature of the companions further, light curves have to
be taken. If the companions are main-sequence stars, they should be
detectable from light variation due to the reflection effect. As an
example we mention HE\,0230$-$4323, an sdB binary with a period of
$0.45\,{\rm d}$ \citep{piii2_edelmann}. \citet{piii2_koen} demonstrated that the
companion is a main-sequence star by measuring the reflection
effect. This star is of particular interest because other, yet
unexplained light variations have been discovered as well.

\end{document}